\input phyzzx
\overfullrule=0pt
\def\psibar{\overline\psi}
\def\Dslash{D\kern-0.15em\raise0.17ex\llap{/}\kern0.15em\relax}
\def\calDslash{{\cal D}%
\kern-0.15em\raise0.17ex\llap{/}\kern0.15em\relax}
\def\sqr#1#2{{\vcenter{\hrule height.#2pt
      \hbox{\vrule width.#2pt height#1pt \kern#1pt
          \vrule width.#2pt}
      \hrule height.#2pt}}}

\def\square{{\mathchoice{\sqr84}{\sqr84}{\sqr{5.0}3}{\sqr{3.5}3}}}
\def\kslash{k\kern-0.026em\raise0.17ex\llap{/}%
          \kern0.026em\relax}
\def\pslash{p\kern-0.026em\raise0.17ex\llap{/}%
          \kern0.026em\relax}
\def\qslash{q\kern-0.026em\raise0.17ex\llap{/}%
          \kern0.026em\relax}
\def\M{{\cal M}}
\def\N{{\cal N}}
\REF\SHA{%
For review, see,
Y. Shamir,
{\it in\/} Lattice 1995, hep-lat/9509023;\hfill\break
H. Neuberger,
hep-lat/9511001, and references therein.}
\REF\NIE{%
H. B. Nielsen and M. Ninomiya,
{\sl Nucl.\ Phys.} {\bf B185} (1981) 20;
{\sl Phys.\ Lett.} {\bf 130B} (1983) 389;\hfill\break
L. H. Karsten,
{\sl Phys.\ Lett.} {\bf 104B} (1981) 315.}
\REF\ADL{%
S. L. Adler, 
{\sl Phys.\ Rev.} {\bf 177} (1969) 2426;\hfill\break
J. Bell and R. Jackiw,
{\sl Nuovo Cim.} {\bf 60A} (1969) 47.}
\REF\BAR{%
W. A. Bardeen,
{\sl Phys.\ Rev.} {\bf 184} (1969) 1848;\hfill\break
C. Bouchiat, J. Iliopolos and Ph.\ Meyer,
{\sl Phys.\ Lett.} {\bf 38B} (1972) 519;\hfill\break
D. J. Gross and R. Jackiw,
{\sl Phys.\ Rev.} {\bf D6} (1972) 477.}
\REF\KAR{%
L. H. Karsten and J. Smit,
{\sl Nucl.\ Phys.} {\bf B183} (1981) 103.}
\REF\KARS{%
L. H. Karsten and J. Smit,
{\sl Phys.\ Lett.} {\bf 85B} (1979) 100.}
\REF\FUJ{%
K. Fujikawa,
{\sl Phys.\ Rev.} {\bf D29} (1984) 285;
{\sl Nucl.\ Phys.} {\bf B428} (1994) 169.\hfill\break
See also, J. Schwinger, {\sl Phys.\ Rev.} {\bf 82} (1951) 664.}
\REF\FUJI{%
K. Fujikawa,
{\sl Phys.\ Rev.\ Lett.} {\bf 42} (1979) 1195;
{\sl Phys.\ Rev.} {\bf D21} (1980) 2848;
{\bf D22} (1980) 1499(E); {\bf D25} (1982) 2584.}
\REF\BARD{%
W. A. Bardeen and B. Zumino,
{\sl Nucl.\ Phys.} {\bf B244} (1984) 421.}
\REF\WES{%
J. Wess and B. Zumino,
{\sl Phys.\ Lett.} {\bf 37B} (1971) 95.}
\REF\FRO{%
However, for real gauge representations, the generalized
Pauli--Villars method provides the action level realization. See,
K. Okuyama and H. Suzuki,
{\sl Phys. Lett.} {\bf B382} (1996) 117; hep-th/9603062,
and references therein.}
\REF\WIL{%
K. Wilson,
{\sl Phys.\ Rev.} {\bf D10} (1974) 2445.}
\REF\WILS{%
K. G. Wilson,
{\it in\/} New Phenomena in Subnuclear Physics, ed.\ by
A. Zichichi (Plenum Press, New York, 1977).}
\REF\KAW{
H. Kawai, R. Nakayama and K. Seo,
{\sl Nucl.\ Phys.} {\bf B189} (1981) 40.}
\REF\THO{
G. 't Hooft,
{\sl Phys.\ Rev.\ Lett.} {\bf 37} (1976) 8;
{\sl Phys.\ Rev.} {\bf D14} (1976) 3432; {\bf D18} (1978) 2199(E).}
\REF\REB{%
C. Rebbi, {\sl Phys.\ Lett.} {\bf B186} (1987) 200.}
\REF\CAM{%
M. Campostrini, G. Curci and A. Pelissetto,
{\sl Phys.\ Lett.} {\bf B193} (1987) 279;\hfill\break
G. T. Bodwin and E. V. Kovacs,
{\sl Phys.\ Lett.} {\bf B193} (1987) 283;\hfill\break
A. Pelissetto, {\sl Ann.\ of Phys.} (NY) {\bf 182} (1988) 177.}

\pubnum={IU-MSTP/16; hep-th/9609054}
\date={November 1996}
\titlepage
\title{Manifestly Gauge Covariant Treatment of Lattice Chiral Fermion}
\author{Hiroshi Suzuki\foot{e-mail: hsuzuki@mito.ipc.ibaraki.ac.jp}}
\address{Department of Physics, Ibaraki University, Mito 310, Japan}
\abstract{%
We propose a lattice formulation of the chiral fermion which
maximally respects the gauge symmetry and simultaneously is free of
the unwanted species doublers. The formulation is based on the
lattice fermion propagator and composite operators, rather than on
the lattice fermion action. The fermionic determinant is defined as
a functional integral of an expectation value of the gauge current
operator with respect to the background gauge field: The gauge
anomaly is characterized as the non-integrability. We perform
some perturbative test to confirm the gauge covariance and an
absence of the doublers. The formulation can be applied rather
straightforwardly to numerical simulations in the quenched
approximation.}
\endpage
The chiral fermion on the lattice has refused its manifestly gauge
invariant treatment~[\SHA]. There even exists the no-go
theorem~[\NIE] for such an endeavor. In the continuum counterpart,
the chiral fermion develops a curious phenomenon, called the quantum
anomaly~[\ADL] or more definitely the gauge anomaly~[\BAR]. It can
be argued that the difficulty in the lattice chiral gauge theory
is a natural consequence of the gauge anomaly.

Suppose that we start with a well regularized fermionic partition
function defined by a manifestly gauge invariant lattice fermion
action. In the continuum limit, the gauge anomaly is a gauge
variation of the partition function. Therefore if the fermion content
is not free of the gauge anomaly, the partition function should not
be gauge invariant---this contradicts with the very gauge invariance
of the formulation. There are two possible resolutions: One is an
appearance of the species doublers which cancel the gauge
anomaly~[\KAR]. Another is a pathology in the continuum limit such
as the non-Lorentz covariance~[\KARS]. The ``trouble'' with the
lattice regularization is that it always regularizes ultraviolet
divergences, even when a gauge invariant regularization should be
impossible due to the gauge anomaly.

The above reasoning suggests that the appearance of the unwanted
doublers is quite natural. However the problem in the conventional
approach is of course that the doublers appear even in the anomaly
free cases. Presumably, the ideal lattice formulation of the chiral
fermion will be the one which distinguishes the anomaly free gauge
representations from the anomalous ones. That unknown gauge invariant
lattice action should have a structure that can be written down, for
example, for the spinor rep.\ of $so(4)$ but not for the fundamental
rep.\ of $su(3)$, because the latter is anomalous. Such an ideal
formulation seems to require a further deeper understanding on the
origin of the quantum anomaly.

In this article, we simply abandon a direct gauge invariant definition
of the fermionic partition function. We take an indirect route.
Nevertheless we attempt to respect the gauge symmetry as much as
possible within a range consistent with the gauge anomaly.

Instead of directly defining the fermion action and the partition
function, we start with the propagator and the gauge current operator
on the lattice. This formulation may be regarded as a first
quantization approach, compared to the conventional ones. The
important fact for us is that although the partition function cannot
be regularized gauge invariantly in general, the gauge current can
always be regularized gauge {\it covariantly\/} even if the gauge
representation is anomalous. This type of regularization scheme in the
continuum theory is known as the covariant regularization~[\FUJ].

In the covariant regularization, fermion loop diagrams are defined as
an expectation value of the gauge current $J^{\mu a}(x)$, in the
presence of the background gauge field. The ultraviolet divergence of
the diagram is then regularized by inserting a gauge invariant dumping
factor into the fermion propagator. In this way, the gauge invariance
associated with all the gauge vertices {\it except\/} that of
$J^{\mu a}(x)$, is preserved. The basic idea is that a possible
breaking of the gauge symmetry due to the anomaly is forced on
{\it the} $J^{\mu a}(x)$ vertex as much as possible. The gauge anomaly
$D_\mu\VEV{J^{\mu a}(x)}$ thus defined has the covariant
form~[\FUJI,\BARD] because of the gauge invariance at external
vertices. On the other hand, a gauge singlet operator such as the
fermion number current is always regularized gauge invariantly.
The scheme thus spoils the Bose symmetry in general but it is
restored when the theory is free of the gauge anomaly. The scheme is
very powerful and applicable to any chiral gauge theories including
the Yukawa couplings.

Once the expectation value of the gauge current is obtained in the
covariant regularization, the fermionic determinant may be defined
as a functional integral of $\VEV{J^{\mu a}(x)}$ with respect to the
background gauge field. However it is obvious that the integration is
possible only when there exists a Bose symmetry among all the gauge
vertices. In other words, the gauge anomaly should satisfy the
Wess--Zumino condition~[\WES] which is a consequence of the
integrability. Since the covariant anomaly breaks the Bose symmetry
and the Wess--Zumino condition~[\FUJI,\BARD], we cannot define the
fermionic determinant from the integration of $\VEV{J^{\mu a}(x)}$,
provided that it has the covariant gauge anomaly.\foot{%
When the gauge group is Abelian, the Wess--Zumino condition is trivial
and gives no constraint. The existence of the covariant anomaly
implies the non-integrability also in this case because,
$0\ne\delta\partial^x_\mu\VEV{J^\mu(x)}/(\delta A_\nu(y))
\ne\partial^x_\mu\delta\VEV{J^\nu(y)}/(\delta A_\mu(x))=0$,
where the left hand side is the $U(1)$ gauge anomaly and the right
hand side means the current is covariantly regularized.}
Therefore, following the covariant regularization, the gauge anomaly
is characterized as the non-integrability of this integration process.
This is a consistent picture because the fermionic determinant
cannot be gauge invariant when the gauge anomaly is present. Our
proposal in this article in spirit can be regarded as a lattice version of the
covariant regularization. We notice that the
covariant regularization itself does not require the action level
realization~[\FRO].

We proceed as follows: In the continuum theory, the propagator of a
massless {\it Dirac\/} fermion is expressed as
$$
\eqalign{
   \VEV{T\psi(x)\psibar(y)}&={-1\over i\Dslash}\delta(x-y)
\cr
   &=i\Dslash
   {\displaystyle1\over\displaystyle g^{\mu\nu}D_\mu D_\nu
           +i[\gamma^\mu,\gamma^\nu]F_{\mu\nu}/4}\delta(x-y),
\cr
}
\eqn\one
$$
where $\Dslash\equiv\gamma^\mu(\partial_\mu+iA_\mu)$ is the vector
type, i.e., non-chiral, covariant derivative and $F_{\mu\nu}$ is the
field strength. In the second line, the denominator has been
rewritten as a {\it second\/} derivative, that may allow a lattice
propagator free of the doubler's massless pole. As the propagator on
the lattice, therefore we take
$$
\eqalign{
   \VEV{T\psi(x)\psibar(y)}&\equiv G(x,y)
\cr
   &\equiv i\Dslash(x)
   {\displaystyle1\over\displaystyle \square(x)
           +[\gamma^\mu,\gamma^\nu][U_{\mu\nu}(x)-1]/(4a^2)}
   \delta(x,y),
\cr
}
\eqn\two
$$
where $\delta(x,y)\equiv\delta_{x,y}/a^4$ and $\Dslash(x)$ is the
standard lattice covariant derivative
$$
   \Dslash(x)\equiv\sum_\mu\gamma^\mu{1\over2a}
   \left[U_\mu(x)e^{a\partial_\mu}
        -e^{-a\partial_\mu}U_\mu^\dagger(x)\right],
\eqn\three
$$
($U_\mu(x)$ is the link variable~[\WIL] and $a$ is the lattice
spacing). To avoid the unwanted massless pole, we define the
covariant lattice d'Alembertian by
$$
   \square(x)\equiv-\sum_\mu{1\over a^2}
   \left[U_\mu(x)e^{a\partial_\mu}
        +e^{-a\partial_\mu}U_\mu^\dagger(x)-2\right].
\eqn\four
$$
For the free theory, this is $2\sum_\mu(1-\cos ak_\mu)/a^2$ in the
momentum space and does not have the doubler's zero at $k_\mu=\pi/a$.
Eq.~\four\ is nothing but the Wilson term~[\WILS] apart from one
extra~$1/a$. In eq.~\two, $U_{\mu\nu}(x)$ is the standard plaquette
variable~[\WIL]: $U_{\mu\nu}(x)\equiv
U_\mu(x)U_\nu(x+a^\mu)U_\mu^\dagger(x+a^\nu)U_\nu^\dagger(x)$.
With the parameterization $U_\mu(x)=\exp[iaA_\mu(x)]$, the lattice
propagator~\two\ obviously reduces to the continuum one~\one\ in the
naive continuum limit. The choice~\two\ is by no means unique and
other definition would work as well.

Using the lattice propagator~\two, we define a fermion bi-linear
operator as\foot{It is equally well easy to define, say, the two
point function of baryon type composite operators. Note that the
present formulation is also applicable to the vector gauge theory
such as QCD.}
$$
   \VEV{\psibar(x)\M\psi(x)}
   \equiv-\tr\M G(x,y)\bigr|_{x=y},
\eqn\seven
$$
where the minus sign is due to the Fermi statistics. The gauge current
of a right handed {\it chiral\/} fermion is simply {\it defined\/}
by taking $\M=T^a\gamma^\mu P_R$ where $P_R\equiv(1+\gamma_5)/2$ is
the chirality projection operator. This is possible because nothing
flips the chirality along the fermion line. When the Yukawa coupling
is involved, this simple recipe using the Dirac propagator does not
work and we will comment on later the generalization.

An important property of the definition~\seven\ is the manifest gauge
covariance. Namely, under the gauge transformation on the link
variable $U_\mu(x)\to V(x)U_\mu(x)V^\dagger(x+a^\mu)$, the
propagator~\two\ is transformed as
$G(x,y)\to V(x)G(x,y)\*V^\dagger(y)$. Consequently the bi-linear
operator~\seven\ transforms
$$
   \VEV{\psibar(x)\M\psi(x)}
   \to\VEV{\psibar(x)V^\dagger(x)\M V(x)\psi(x)},
\eqn\eight
$$
which means that the composite operator has a definite transformation
property under the gauge transformation on the external gauge field.
Note that the covariance holds for a finite lattice spacing as well as
the continuum limit $a\to0$. In particular, a gauge {\it singlet\/}
operator, for which $\M$ commutes with the gauge generator, is
regularized {\it gauge invariantly}. In the continuum limit, the gauge
anomaly should have the gauge covariant form provided that the limit
is not pathological.

Since we are not assuming the underlying fermion action in the
present formulation, various symmetric properties are unfortunately
not manifest. Nevertheless the gauge covariance~\eight\ is powerful
enough to derive the Ward identity associated with external gauge
vertices. The vertex function, being a gauge current type operator
$\M=T^a\gamma^\mu\N$ inserted, is defined by
$$
\eqalign{
   \VEV{\psibar(x)T^a\gamma^\mu\N\psi(x)}
   &\equiv\sum_{n=1}^\infty{1\over n!}\prod_{j=1}^n
   \biggl[
   a^4\sum_{x_j,\mu_j,a_j}A_{\mu_j}^{a_j}(x_j)
   \int_{-\pi/a}^{\pi/a}
   {d^4p_j\over(2\pi)^4}e^{ip_j(x-x_j)}e^{ia{p_j}_{\mu_j}/2}
   \biggr]
\cr
\noalign{\vskip2pt}
   &\qquad\qquad\qquad\times
 \Gamma_\N^{\mu\mu_1\cdots\mu_naa_1\cdots a_n}(p_1,p_2,\cdots,p_n),
}
\eqn\nine
$$
where the term independent of the gauge field ($n=0$) identically
vanishes. For example, when the constant matrix $\N$ in~\nine\
commutes with the gauge generator, we find
$$
   p_\nu\lim_{a\to0}\Gamma_\N^{\mu\nu ab}(p)=0,
   \quad
   -ip_\nu\lim_{a\to0}\Gamma_\N^{\mu\nu\rho abc}(p,q)
   -f^{bad}\lim_{a\to0}\Gamma_\N^{\mu\rho dc}(p)=0,
\eqn\ten
$$
and higher point identities, by examining a variation of the both
sides of~\nine\ under an infinitesimal gauge transformation. The
second relation is nothing but the covariant convergence of the gauge
current at one of external vertices in the three point function.
Note however that the gauge covariance~\eight\ itself does {\it not\/}
imply,
$$
   i(p_\mu+q_\mu)\lim_{a\to0}\Gamma_\N^{\mu\nu\rho abc}(p,q)
   -f^{abd}\lim_{a\to0}\Gamma_\N^{\nu\rho dc}(p)=0.
\eqn\eleven
$$
If this relation would hold, it implies the covariant convergence of
arbitrary current operators and contradicts with possible anomalies.
The crucial point in this formulation is that vertices associated
with the external gauge field and the vertex of the composite current
operator are differently treated. Thus in general
$\lim_{a\to0}\Gamma_\N^{\mu\nu\rho abc}$ is not symmetric under an
exchange $\mu\leftrightarrow\nu$ and $a\leftrightarrow b$. It is this
breaking of the Bose symmetry which allows the manifest gauge
covariance of the formulation. However, as was already noted, the Bose
symmetry will be restored in the continuum limit when the theory is
free of the gauge anomaly.

To verify the above properties and that the unwanted doublers are
really absent in the present formulation, we explicitly evaluated some
of the vertex functions in the continuum limit. After a somewhat
lengthly calculation using the technique in~[\KAW], we find that the
two point function is given by
$$
\eqalign{
   &\lim_{a\to0}\Gamma_\N^{\mu\nu ab}(p)
\cr
   &=-{1\over48\pi^2}\tr T^aT^b\gamma^\mu\N
                        (\pslash p^\nu-\gamma^\nu p^2)
   \left[\log{4\pi\over-a^2p^2}
         -\gamma+{5\over3}+4\pi^2(J-{5\over24}K)\right],
\cr
}
\eqn\twelve
$$
where $J=0.0465\cdots$ and $K=0.309\cdots$ are numerical
constants~[\KAW]. The Lorentz covariance is restored and there is no
non-local divergence in~\twelve. Also the quadratically divergent
terms, which are proportional to $g^{\mu\nu}/a^2$, are canceled out,
as the Ward identity~\ten\ and the hypercubic symmetry indicate. By
taking $\N=P_R$, eq.~\twelve\ gives the vacuum polarization tensor
of a right handed chiral fermion:
$$
   \Pi^{\mu\nu ab}(p)
   =-{1\over24\pi^2}\tr T^aT^b(p^\mu p^\nu-g^{\mu\nu}p^2)
   \left[\log{4\pi\over-a^2p^2}
         -\gamma+{5\over3}+4\pi^2(J-{5\over24}K)\right].
\eqn\fourteen
$$
It is transverse, as is constrained by the Ward identity~\ten\
{\it and\/} the logarithmic divergence has the correct coefficient as
a {\it single\/} chiral fermion. Thus we see that the formulation in
fact respects the gauge covariance and simultaneously is free of the
species doublers at least in the perturbative treatment.

For the three point function, we computed the divergence of the vector
and the axial gauge currents:
$$
\eqalign{
   &i(p_\mu+q_\mu)\lim_{a\to0}
   \Gamma_{1}^{\mu\nu\rho abc}(p,q)
   =0,
\cr
   &i(p_\mu+q_\mu)\lim_{a\to0}
   \Gamma_{\gamma_5}^{\mu\nu\rho abc}(p,q)
   ={i\over4\pi^2}\tr T^a\{T^b,T^c\}
   \varepsilon^{\nu\rho\alpha\beta}p_\alpha q_\beta.
\cr
}
\eqn\fifteen
$$
Both relations are consistent with the Ward identity and actually
the first of~\fifteen\ may be derived solely from eq.~\ten\ and a
general argument. The second relation should be interpreted as the
covariant anomaly because of the underlying gauge covariance: It has
the unique covariantized form
$$
\eqalign{
   &D_\mu\lim_{a\to0}
   \VEV{\psibar(x)T^a\gamma^\mu\psi(x)}
   =0,
\cr
   &D_\mu\lim_{a\to0}
   \VEV{\psibar(x)T^a\gamma^\mu\gamma_5\psi(x)}
   ={i\over16\pi^2}\varepsilon^{\mu\nu\rho\sigma}
   \tr T^aF_{\mu\nu}F_{\rho\sigma}.
\cr
}
\eqn\seventeen
$$
Therefore the gauge anomaly of a chiral fermion (note that $P_R$ is
inserted in the gauge current) has the covariant form with the correct
coefficient. We also find the correct fermion number anomaly~[\THO]
of a single chiral fermion by substituting $T^a\to 1$ in~\seventeen.

We have observed that, besides the manifest gauge covariance~\eight,
the present formulation possesses many desired features at least in
the perturbative treatment. At this point the reader might be
wondering if the present formulation is equivalent to a non-local
fermion action
$$
   S=a^4\sum_x\psibar(x)\left\{
   \square(x)+[\gamma^\mu,\gamma^\nu][U_{\mu\nu}(x)-1]/(4a^2)
   \right\}{-1\over i\Dslash(x)}P_R\psi(x),
\eqn\eighteen
$$
because it obviously corresponds to the propagator~\two, hence the
non-locality leads to some pathology. This interpretation is
{\it not\/} correct. If our formulation is simply based on the
{\it action\/}~\eighteen, the gauge current would be defined by
$\VEV{\psibar(x)T^a\gamma^\mu\psi(x)}_{\rm BS}
 \equiv-\delta
 \log\int\prod_yd\psibar(y)d\psi(y)e^S/(\delta A_\mu^a(x))$
and the definition obviously respects the Bose symmetry among all the
gauge vertices. As a consequence, we have the consistent form of gauge
anomaly in the continuum limit which contradicts with the manifest
gauge invariance of~\eighteen. As was already argued, we then expect
the doublers or, a pathology such as a breaking of the Lorentz
covariance. Our formulation based on the prescription~\seven\ with
$\M=T^a\gamma^\mu P_R$, on the other hand, explicitly
spoils the Bose symmetry but instead respects the maximal background
gauge covariance. The possible gauge anomaly has the covariant form.
Therefore two approaches are completely different {\it even in the
continuum limit}.

For simplicity of the presentation, we have neglected the possible
Yukawa couplings up to now, which is important in realistic
chiral gauge theories. The generalization of \two\ and~\seven\ is
however straightforward. In the continuum theory, the covariant
derivative is generalized as
$\calDslash\equiv\Dslash_RP_R+\Dslash_LP_L-iG\phi_RP_R-iG\phi_LP_L$,
with obvious notations. The expression~\one\ is therefore replaced
by
$-1/(i\calDslash)=i\calDslash^\dagger/(\calDslash\calDslash^\dagger)$
and the following steps are almost identical. The lattice
d'Alembertian~\four\ may be used for the right handed and the left
handed components respectively.

From the above analyses, the present proposal seems to provide a gauge
covariant (or invariant for a gauge singlet operator) definition of
composite operators without the unwanted doublers. We then have to
integrate the gauge current expectation value to construct
the fermionic determinant. A cancellation of the gauge anomaly is
the integrability condition in the continuum limit, as was already
noted. However this fact is not so useful practically because the
analytical integration is a formidable task and, the continuum limit
is never reached in numerical simulations. Clearly we ought to study
the integrability with a {\it finite\/} lattice spacing (and
associated modifications of \two\ and~\seven, if necessary) for
setting up a non-perturbative framework. This analysis is in
progress and will be reported elsewhere. Here we simply note that
what is needed in the Metropolis simulation is not the fermionic
determinant itself but the {\it difference\/} of the determinant
between two gauge field configurations. This is the lattice analogue
of the gauge current expectation value.

However the integration is not necessary at all if one is contented
with the {\it quenched\/} approximation. The application to numerical
simulations is straightforward once having the lattice fermion
propagator such as~\two. We therefore believe that our proposal, even
in the present form, has a range of practical application at least
within the quenched approximation.

The author would like to thank K.~Fujikawa, S.~Kanno and Y. Kikukawa
for discussions. This work is supported in part by the Ministry of
Education Grant-in-Aid for Scientific Research, Nos.~08240207,
08640347 and~07304029.
\medskip
\noindent
{\bf Note added:}\quad
After this paper was accepted for publication, the author
was aware of a similar proposal had already been made~[\REB].
However the point that we do not assume the underlying non-local action
is the crucial difference. Our consideration also explains why the
correct axial anomaly evaluated from a composite current operator
definition~[\REB] and the correct vacuum polarization tensor are not
reproduced in the corresponding non-local action calculations~[\CAM].

\refout
\bye